\documentclass[fleqn,usenatbib]{mnras}

\usepackage{newtxtext,newtxmath}

\usepackage[T1]{fontenc}

\DeclareRobustCommand{\VAN}[3]{#2}
\let\VANthebibliography\thebibliography
\def\thebibliography{\DeclareRobustCommand{\VAN}[3]{##3}\VANthebibliography}
\usepackage{xcolor}
\usepackage{soul}

\usepackage{graphicx}	
\usepackage{amsmath}	

\title[PKS~0735+178: blazar-neutrino association]{Dissecting the broadband emission from $\gamma$-ray blazar PKS~0735+178 in search of neutrinos}

\author[Authors et al.]{
Raj Prince,$^{1}$\thanks{E-mail: raj@cft.edu.pl}
Saikat Das,$^{2}$
Nayantara Gupta,$^{3}$
Pratik Majumdar, $^{4}$ and
Bo{\.z}ena Czerny$^{1}$
\\
$^{1}$Center for Theoretical Physics, Polish Academy of Sciences, Al.Lotnikow 32/46, 02-668, Warsaw, Poland\\
$^{2}$Center for Gravitational Physics and Quantum Information, Yukawa Institute for Theoretical Physics, Kyoto University, Kyoto 606-8502, Japan\\
$^{3}$Astronomy \& Astrophysics Group, Raman Research Institute, C.V. Raman Avenue, Sadashivanagar, Bangalore 560080, Karnataka, India\\
$^{4}$Saha Institute of Nuclear Physics, a CI of Homi Bhabha National Institute, Kolkata, West Bengal 700064, India\\
}

\date{Accepted XXX. Received YYY; in original form ZZZ}

\pubyear{2022}

\begin{document}
\label{firstpage}
\pagerange{\pageref{firstpage}--\pageref{lastpage}}
\maketitle

\begin{abstract}
The origin of the diffuse flux of TeV-PeV astrophysical neutrinos is still unknown. The $\gamma$-ray blazar PKS 0735+178, located outside the 90\% localization region at 2.2 deg from the best-fit IC-211208A event,
 was found to be flaring across all wavebands.
In addition to leptonic synchrotron (SYN) and synchrotron self-Compton (SSC) emission, we invoke photohadronic ($p\gamma$) interactions inside the jet to model the spectral energy distribution (SED) and neutrino emission.
 We analyze the 100 days $\gamma$-ray and X-ray data and 10 days around the neutrino event is chosen to generate the broadband SED. The temporal light curve indicates that the source was in a high state in optical, UV, $\gamma$-ray, and X-ray frequencies during the neutrino detection epoch. 
In the one-zone lepto-hadronic model, the SSC photons do not provide enough seed photons for $p\gamma$ interactions to explain the neutrino event. However, including an external photon field yields a neutrino event rate of 0.12 in 100 days, for the IceCube detector, using physically motivated values of the magnetic field, an external photon field peaking at optical wavelength, and other jet parameters. 
The radiation from secondary electrons at X-ray energies severely constrains the neutrino flux to a lower value than found in previous studies. Moreover, the flux of high-energy $\gamma$-rays at GeV energies from the decay of neutral pions is subdominant at the high-energy peak of the SED, suggesting a higher correlation of neutrinos flux with X-ray flux is plausible.

\end{abstract}




\section{Introduction}

The origin of the diffuse flux of astrophysical neutrinos from $\sim$10 TeV to a few PeV, detected by IceCube Observatory, is a relatively recent problem in astroparticle physics \citep{Aartsen_2013, Aartsen_2014}. After ten years of their discovery, their sources are still concealed from the eyes of multiwavelength telescopes. Since neutrinos are produced only in hadronic processes, they are smoking gun evidence of cosmic-ray (CR) acceleration inside the source \citep[see for eg.,][]{Aartsen_2018}. CRs consist mainly of protons and
nuclei and can interact with the ambient matter and radiation fields at the production site and/or during their propagation to Earth. Astrophysical neutrinos are produced along with high-energy photons via photohadronic ($p\gamma$) or hadronuclear ($pp$) interactions in the source emission region. The magnetic rigidity at tens of PeV energies is insufficient for retaining directional signatures of cosmic rays over cosmological distances. The high-energy universe is opaque to photons beyond a few hundred TeV due to $e^+e^-$ pair production with cosmic background photons. Neutrinos are ideal messengers of cosmic-ray acceleration because they are weakly interacting, and thus unattenuated by intervening matter or radiation and undeflected by magnetic fields. Thus the discovery of the neutrino emitters will unveil the sources of extragalactic cosmic rays, and even the ultrahigh-energy cosmic rays \citep[UHECRs; $E>10^{17}$ eV, see, e.g.,][for a recent review]{Anchordoqui:2018qom, AlvesBatista:2019tlv}.

The real-time alert system at the IceCube neutrino Observatory, located in the South Pole ice, allows rapid follow-up by multiwavelength detectors to search for the source of a high-energy muon-neutrino event \citep{IceCube:2016cqr}. Neutrinos of astrophysical origin at $\gtrsim1$ TeV and declination $\delta \gtrsim 5^\circ$ can be efficiently discriminated from the atmospheric muon contamination with an angular resolution of $<1^\circ$ for track-like events. Active Galactic Nuclei (AGNs), powered by accretion onto a supermassive black hole, are considered prominent candidates of IceCube astrophysical neutrino signal. 
 The association of the neutrino event IC-170922A with a flaring Fermi-LAT blazar TXS~0506+056, at 3$\sigma$ statistical significance led to the first spatial and temporal correlation of a high-energy $\nu_\mu$ event with a flaring $\gamma$-ray blazar, detected by Fermi-LAT, and followed up by several detectors in an intensive multiwavelength campaign \citep{IceCube:2018dnn, IceCube:2018cha}. Several other neutrino events are identified thereafter having positional coincidence with blazars but with lower statistical significance \citep{Franckowiak:2020qrq, Giommi:2020hbx}.

Blazars are a class of AGN with their collimated beam of outflow pointing toward the observer's line of sight (\citealt{1995PASP..107..803U}). These powerful luminous jets are relativistic, extending from pc to kpc scale distances. Blazar jets can provide a suitable environment for cosmic-ray acceleration and neutrino production \citep{IceCube:2022zbd}. The blazars with weak or no emission lines are called BL Lacs; they are usually highly variable sources. Based on the frequency of their synchrotron emission peaks ($\nu_{peak}$) in broadband SED, BL Lacs are classified into low energy-peaked (LBL) ($\nu_{peak}\approx 10^{13-14}$ Hz), intermediate energy-peaked (IBL) ($\nu_{peak}\approx 10^{15-16}$ Hz) and high energy-peaked (HBL) ($\nu_{peak}\approx 10^{17-18}$ Hz) sources. More recently, the LBLs with peak frequency above $10^{13.5}$ Hz have been named intermediate or high energy-peaked BL Lacs (IBL/HBL) \citep{giommi2021astrophysical}. Earlier analyses have shown, however, that a maximum of 30\% of the cumulative neutrino background can originate from blazars, depending on the scaling relation between $\gamma$-rays and neutrinos \citep{IceCube:2016qvd}.

It is possible that along with electrons, protons are also accelerated to very high energy inside the jet. 
In the leptonic model, the broadband spectral energy distribution can be well explained by the synchrotron and inverse-Compton emission. It has been noted earlier that in some cases,
both leptonic and hadronic models can explain the broadband SED, and hence a clear detection of neutrino events is required where the 
 hadronic modeling is preferred (\citealt{2001APh....15..121M, Bottcher2013, 10.1093/mnras/stu2364}).
 In hadronic interactions, the shock-accelerated protons or heavy nuclei interact with the cold protons in the ambient medium (p-p interaction) or low-energy photons (p-$\gamma$ interaction) and produce the high-energy gamma-rays as well as neutrinos \citep{Bottcher2013, Reimer:2018vvw}. Although both $\gamma$-rays and neutrinos maintain their directionality, due to their non-interacting nature, high-energy neutrinos are the best candidates to probe the acceleration site of cosmic rays and their maximum energy \citep{Sikora_1987, Eichler_1979}.

PKS~0735+178 is an IHBL located at a redshift of $z=0.45\pm0.06$ as determined from the detection of its host galaxy \citep{2012A&A...547A...1N}. The redshift value of 0.65 has also been suggested in a more recent study by \citet{2021ATel15132....1F}, but we used 0.45 in this work. During its largest flaring state in December 2021, it was found to be associated with neutrino events detected by IceCube \citep{2021GCN.31191....1I}, Baikal \citep{2021ATel15112....1D}, Baksan \citep{2021ATel15143....1P}, and KM3NeT \citep{2022ATel15290....1F}. We have analyzed the observational data from {\it Swift-XRT}, {\it UVOT}, {\it NuSTAR} and {\it Fermi-LAT} to build the multi-wavelength spectral energy distribution (SED) during the flaring state. We initially modeled the SED with a leptonic model, which can explain the SED satisfactorily, and subsequently with a lepto-hadronic model to determine whether PKS~0735+178 could be the source of astrophysical neutrinos. An external radiation field, which may originate from the Broad Line Region (BLR), is needed to enhance the neutrino flux. 
We have discussed the neutrino event detected by the IceCube collaboration in section 2 and the analysis of the UV, X-ray, and $\gamma$-ray data in section 3. Our results are presented in section 4, they are discussed in section 5, and the conclusions from our results are given in section 6.

\section{Neutrino event and its follow-up}
A GCN circular 31191 was posted confirming the detection of a high-energy neutrino candidate with a track-like event (IceCube-211208A; \citealt{2021GCN.31191....1I}). In IceCube, the detection was recorded on 2021-12-08 at 20:02:51.1 UT with a moderate probability of being of astrophysical origin. Fermi-LAT reported a significant detection in gamma-ray from the blazar PKS 0735+178 (4FGL J0738.1+1742) which was located at 
 $\sim$2.2 degrees away from the event constructed location by the IceCube \citep{2021GCN.31194....1G}, which triggered a massive search of its counterparts in different wavebands across the globe. Soon after, many Atels were reported (\citealt{2021ATel15021....1S, 2021ATel15113....1F, 2021ATel15108....1H, 2021ATel15105....1K, 2021ATel15143....1P, 2021ATel15136....1L, 2021ATel15148....1C}). The first broadband confirmation came from the optical MASTER telescope which reported a brightening in the optical band (\citealt{2021ATel15098....1Z}) with magnitude 14.1 and soon after the source was observed in a high flux state in $\gamma$-ray (\citealt{2021ATel15099....1G}), X-ray (\citealt{2021ATel15102....1S}), and radio bands (\citealt{2021ATel15105....1K}). The simultaneous brightening in almost all the electromagnetic regimes during the IceCube detection suggests that this blazar could be the possible source of the neutrino.
This motivates us to do a broadband study to understand the physical mechanism responsible for the broadband emission and explore the neutrino connection with the blazar.

\section{Broadband observations and data analysis }
We collected the archival data that is available for this source during this important event and provided a detailed broadband study including the temporal and spectral properties.
\subsection*{Fermi-LAT:}
We analyzed the Pass8 Fermi-LAT dataset within energy 100 MeV to 350 GeV of the source 4FGL J0738.1+1742. We use the \texttt{fermipy} package developed by \cite{2017ICRC...35..824W} for analyzing the \textit{Fermi}-LAT data. A circular $10^{\circ}$ region of interest (ROI) was selected around the source for the photons extraction. The events with zenith angle $\theta <  90^\circ$ have been rejected to avoid contamination from the earth's limb $\gamma$-ray. The filter "gtmktime"(DATA\_QUAL>0 \&\& LAT\_CONFIG==1) is used to select the good time intervals (GTIs). While analyzing the data both the front and back-type events were considered. The above analysis has been performed with the latest instrument response function (IRF) ``P8R3\_SOURCE\_V3", isotropic background model ``iso\_P8R3\_SOURCE\_V3\_v01'' and the galactic diffuse emission model ``gll\_iem\_v07''. These models can be found on Fermi Science Support Center \footnote{\url{https://fermi.gsfc.nasa.gov/ssc/data/access/lat/BackgroundModels.html}}.

\begin{figure}
    \includegraphics[width=0.49\textwidth]{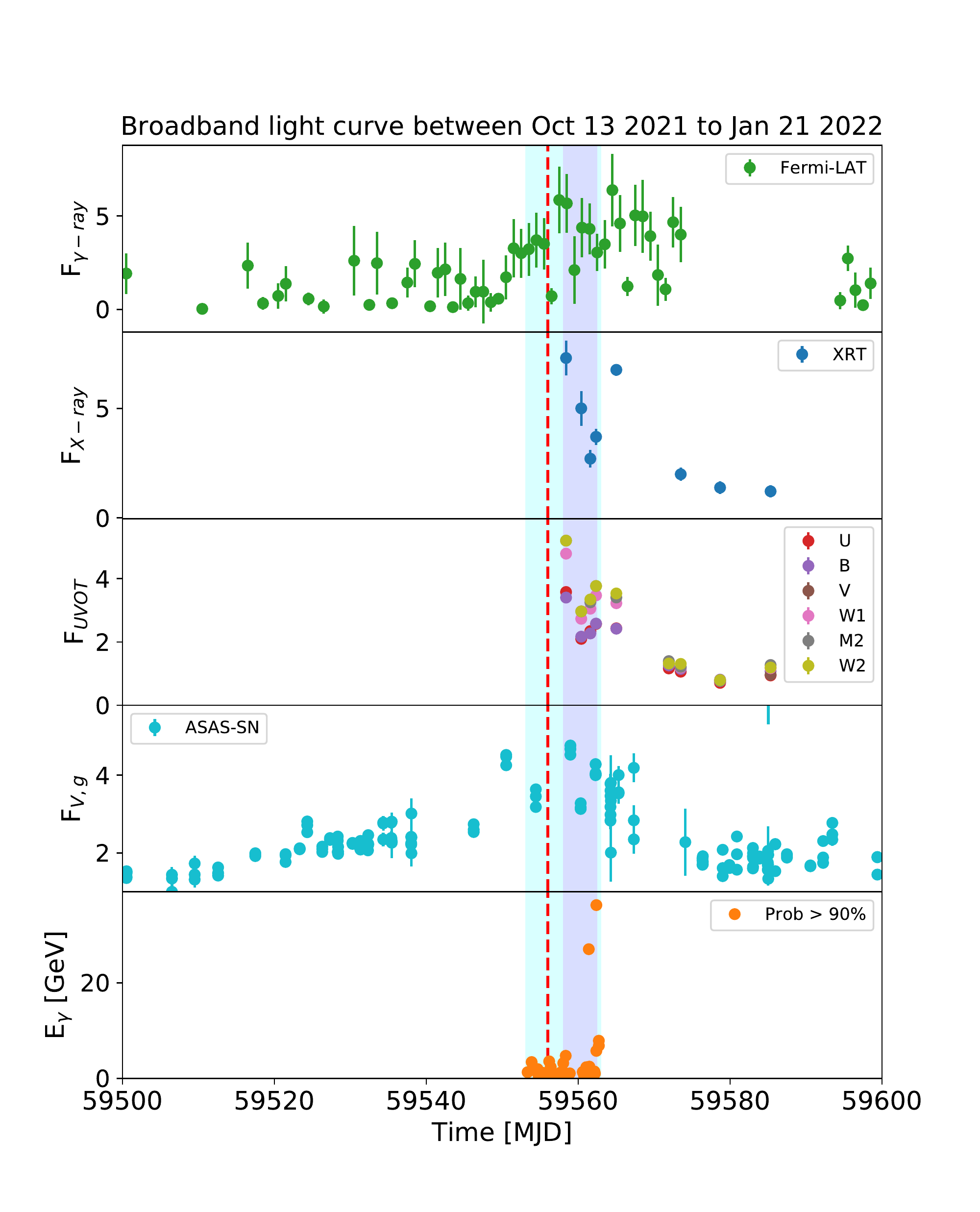}
    \caption{The broadband lightcurve for PKS 0735+17. The shaded region is chosen for SED modeling. The $\gamma$-ray data are in units of 10$^{-7}$ ph cm$^{-2}$ s$^{-1}$, XRT are in 10$^{-12}$ erg cm$^{-2}$ s$^{-1}$, UVOT are in 10$^{-11}$ erg cm$^{-2}$ s$^{-1}$, and ASAS-SN are in units of mJy. The period marked in cyan color is used for SED modeling (MJD 59553-59563). The light cyan color represents the period for $\gamma$-ray SED and the blue color is for X-ray and optical-UV SED.}
    \label{MWL}
\end{figure}

 The \texttt{Likelihood} analysis provides the TS value of each source within the ROI and for further analysis we reject the sources with low TS (TS < 10). Then
we produce the light curve of the interested source keeping its parameters free and by fixing the model parameters of all other sources present within the 10 degrees of the region of interest (ROI).
The upper panel of Figure \ref{MWL} shows the weekly binned $\gamma$-ray light curve. The source was analyzed from Oct 2021 to Jan 2022.

To generate the spectral energy distribution (SED) of the sources for modeling, we keep the parameters of the galactic and isotropic diffuse components free, and initially,  the parameters of all the other sources present within the ROI are kept fixed. We also reject the sources with a lower TS (TS < 10) value.
We calculate flux upper limits in energy bins if the source TS is found to be less than 9. 

The photon index of the target was left free to vary (along with all the other parameters) but that of the other sources was kept fixed. For the background sources that satisfied the TS$\geq$10 limit, only the normalization or prefactor was left free to vary (depending on the model being Power law or LogParabola) whereas their spectral indices were kept fixed. The effect of energy dispersion was acknowledged during the analysis except for 'galdiff' and 'isodiff' for which the effect of energy dispersion was ignored.

\subsection*{Swift-XRT:}
Soon after detecting the neutrino event and the high source activity in $\gamma$-ray, the source was proposed for monitoring with \texttt{Swift} telescope in X-ray and optical-UV wavebands. We have collected and analyzed the long-term data and also focused on the period close to the neutrino event. The analysis is done following the standard Swift-XRT procedure. A source and background region of 15$"$ and 30$"$ is chosen around the source and away from the source to extract their corresponding spectrum. The ancillary response file (ARFs) corresponding to each observation is created using the tool \texttt{XRTMKARF} and the corresponding re-distribution matrix file (RMFs) is obtained from the latest version of the CALDB. The RMFs and ARFs file along with the source and background spectrum is loaded into the tool \texttt{GRPPHA} to combine and group them. The final output spectrum of the \texttt{GRPPHA} is binned with 20 counts per bin to achieve the minimum of 20 counts in each bin. The final spectra are loaded into \texttt{XSPEC} and modeled with a simple power law spectral model. The X-ray photons in the soft energy band are accounted for the galactic correction using the galactic column density, n$_H$ = 4.5$\times$10$^{20}$ cm$^{-2}$ (\citealt{2013MNRAS.431..394W}). 

\subsection*{Swift-UVOT:}
The advantage of having \texttt{Swift} telescope is that it can provide simultaneous observations in optical-UV along with the X-ray. The Swift has an ultraviolet optical telescope (UVOT) onboard with six filters namely, UVV, UVU, UVB, UVW1, UVM2, and UVW2 where the first three are optical filters and the latter three are ultraviolet. The raw data were analyzed following the standard procedure where the source and background region were selected with a circular region of radius 5$"$ and 15$"$ around the source and away from the source location, respectively. The magnitude was extracted using the tool \texttt{UVOTSOURCE} from each filter and later was corrected for galactic extinction using the reddening, E(B-V) = 0.0298 (\citealt{2011ApJ...737..103S}) and extinction ratio corresponding to each wavelength, A$_\lambda$/E(B-V) from \citealt{2006A&A...456..911G}. Further, the corrected magnitude was converted to fluxes using the zero point (\citealt{2011AIPC.1358..373B}) and the conversion factors (\citealt{2008MNRAS.383..627P}).

\subsection*{NuSTAR:}
Following the IceCube event the source was proposed for ToO observation in NuSTAR and two observations were done on 11 December 2021 and 13 December 2021 with $\sim$22 \& $\sim$21 ks exposure  \citep{2021ATel15113....1F}. We have analyzed these two observations using the standard NuSTAR data analysis tool \texttt{NuSTARDAS} 1.9.2 provided by \texttt{HEASOFT}. \texttt{Nupipeline} was run on both observations to produce the cleaned event file. A circular source and background region were chosen of radius 20$"$ and 40$"$, respectively. A tool called \texttt{nuproduct} was run to create the source and background spectra along with corresponding \texttt{rmf} and \texttt{arf} files. Further, the spectra were processed in \texttt{xspec}. 

\section{Results}
\subsection{Broadband Light curves}
\begin{figure}
\includegraphics[scale=0.45]{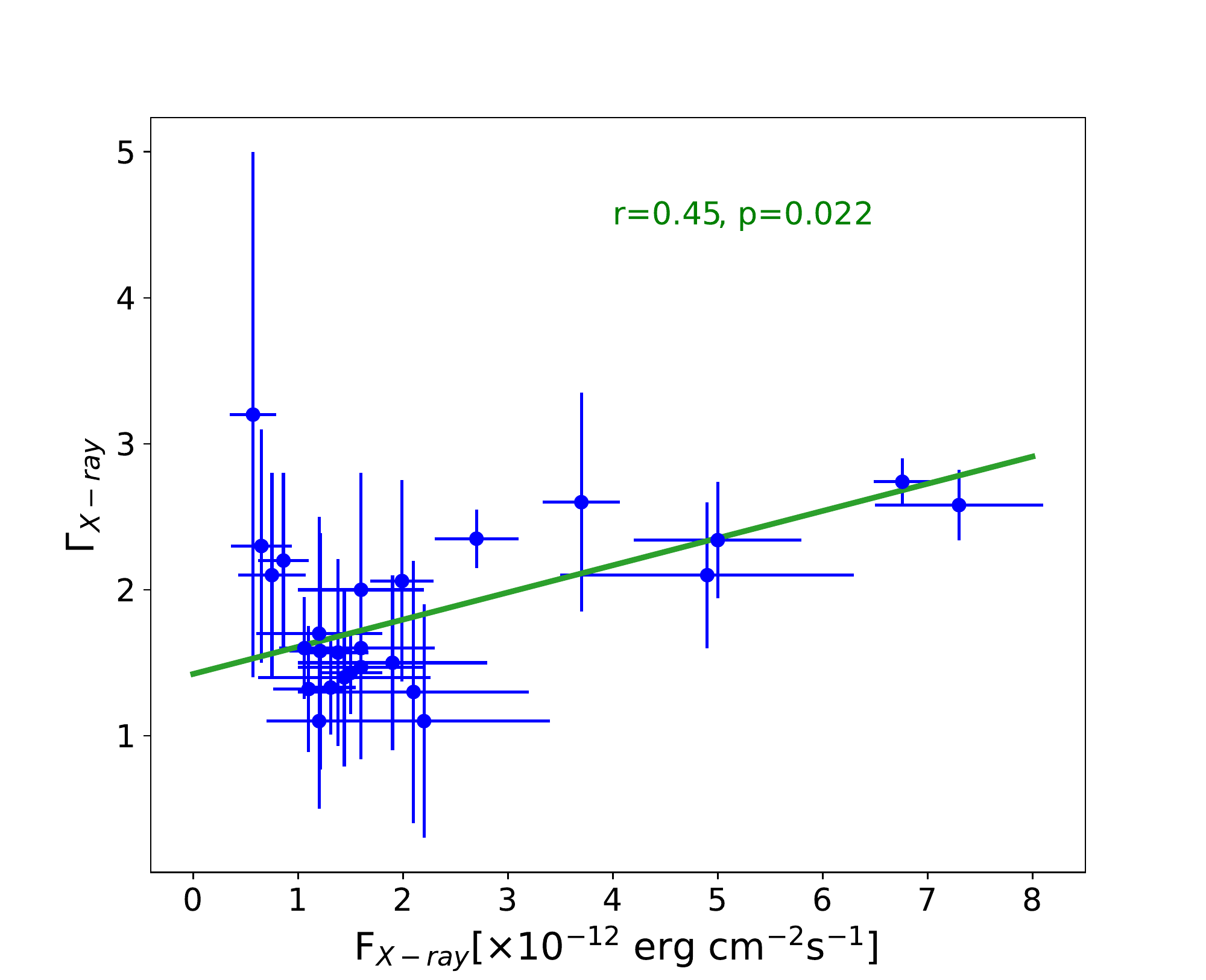}

\caption{ The Flux-Index variation of X-ray emission during the long period. The estimated correlation coefficient('r') is 45$\%$ with a null hypothesis probability('p') of 0.02, suggesting a "softer-when-brighter" trend which resembles the LBL-type source \citep{giommi2021astrophysical}.}
\label{flux-index}
\end{figure}

After the IceCube alert, the source was rigorously monitored across the wavebands. We have shown the temporal behavior of the source during the event and a few weeks before and after the events to have a clear understanding of its variability. 
 Figure \ref{MWL} presents the broadband light curve over 100 days starting from 1.5 months before the neutrino event (red dashed vertical line). Before and after the IceCube event the gamma-ray flux is low and the detection is not significant in some epochs (TS$<$9; those data points are removed from the light curve).
We noticed that during the neutrino event, the source was rising in $\gamma$-rays where the flux rises from $\sim$2$\times$10$^{-7}$ ph cm$^{-2}$ s$^{-1}$ to $\sim$6.3$\times$10$^{-7}$ ph cm$^{-2}$ s$^{-1}$ within a day. Then it goes down again reaching the flux level $\sim$2$\times$10$^{-7}$ ph cm$^{-2}$ s$^{-1}$ within 2 days and again started rising. This confirms a short-term variability in the $\gamma$-ray light curve. However, the variability does not coincide exactly with the neutrino event (vertical red line in Figure \ref{MWL}). Similar behavior was also seen in the X-ray band where we observed that the source was in a high state (as the later observation suggests) but the observation at the exact IceCube event date was missed in Swift. Just after the event, the X-ray observation was done and it showed a clear fast-flux variability (before the events within a few months there were no observations made in X-ray by Swift-XRT). The flux goes down by a factor of three from 7.3$\times$10$^{-12}$ to 2.7$\times$10$^{-12}$ ergs cm$^{-2}$ s$^{-1}$ just within 3 days after the event and again rises to the high flux state to 6.76$\times$10$^{-12}$ ergs cm$^{-2}$ s$^{-1}$ within 3 days and then went down to a low flux state. During this variability, we noticed that the photon spectral index does not change much but we see a slight trend of the "softer-when-brighter". This trend is more clear when plotting all the observations together in Figure \ref{flux-index}. The trend is that of an LBL-type blazar as also seen in \citet{2021MNRAS.507.5690G}.

Using the 1-day binned $\gamma$-ray light curve, we have estimated the flux variability time and the size of the emission region. For that, we followed the expression,
\begin{equation}
    F(t_2) = F(t_1) \times 2^{(t_2 - t_1)/T_d}
\end{equation}
Here $F(t_1$) and $F(t_2$) are the observed fluxes at time $t_1$ and $t_2$, respectively and $T_d$ is the flux doubling/halving time scale which means within this time the source flux increases or decreases by a factor of 2. The fastest doubling/halving time in $\gamma$-ray was found to be 0.50$\pm$0.08 days or 12 hours. 
The hour scale variability is very common in blazars suggesting a compact emitting region. 
The size of the emission region (R) can be estimated by using the equation,
\begin{equation} \label{eq:3}
    R' = c \delta t_{\rm var}/(1+z),
\end{equation}
 where $R'$ is estimated to be 2.68$\times$10$^{16}$cm, using the $\delta$=30 and $t_{\rm var}=$ 0.5 days, and redshift = 0.45.

Simultaneous to $\gamma$-ray and X-ray this source appears to be in a high state also in optical-UV. In UVOT, we do not have any optical-UV observation on the exact date of neutrino detection but the observation done after the event suggests a brightening. We also have access to the long-term ASAS-SN  optical data as shown in panel 4, which confirms the optical brightening during the flaring and the IceCube event. In the lower panel, we show the detection of high-energy photons with probability $>$99$\%$. Two high-energy photons of energy above 30 GeV were detected after the IceCube event.




\subsection{Multiwavelength SED and neutrino production}

\begin{figure*}
\includegraphics[width=0.49\textwidth]{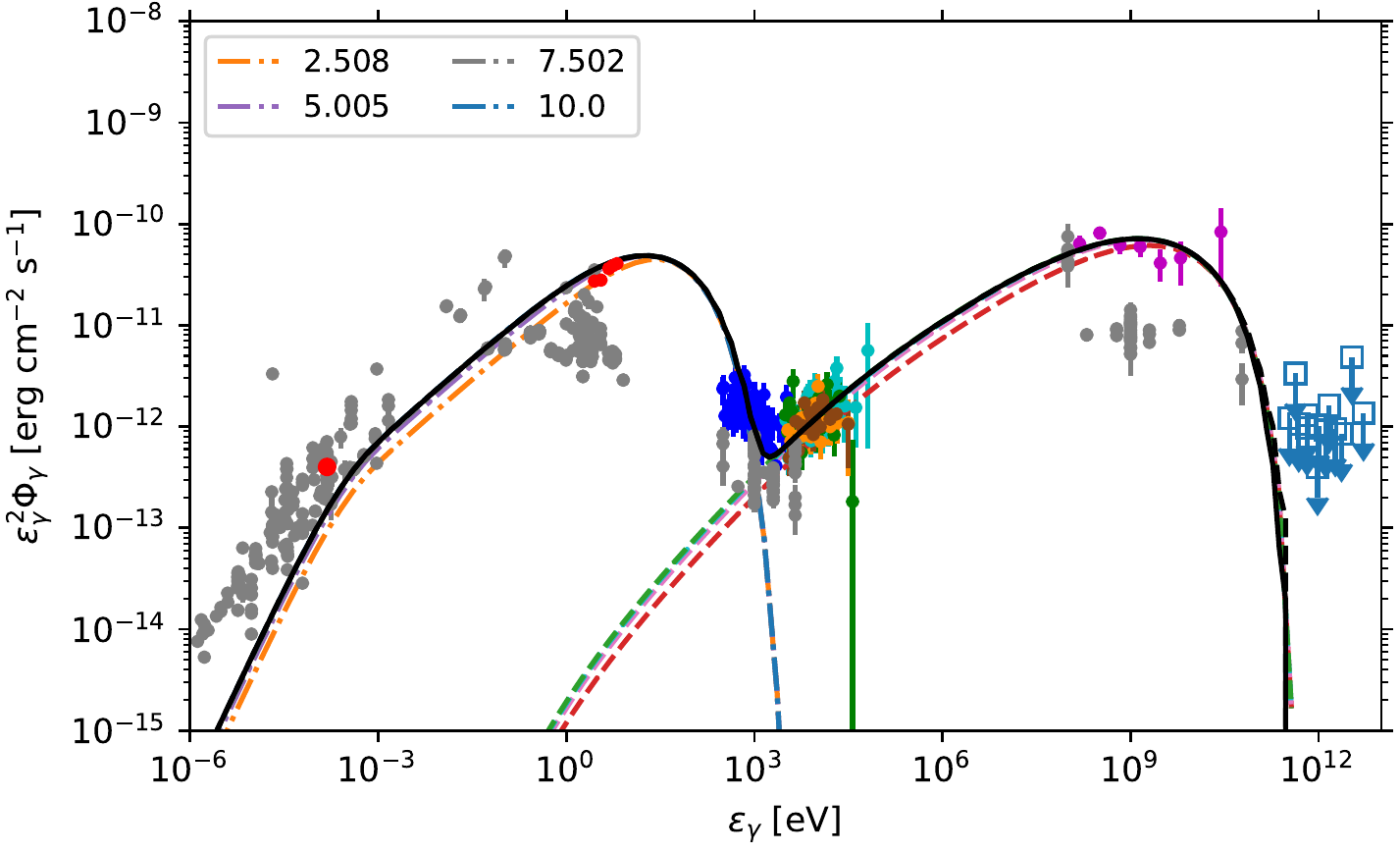}
\includegraphics[width=0.49\textwidth]{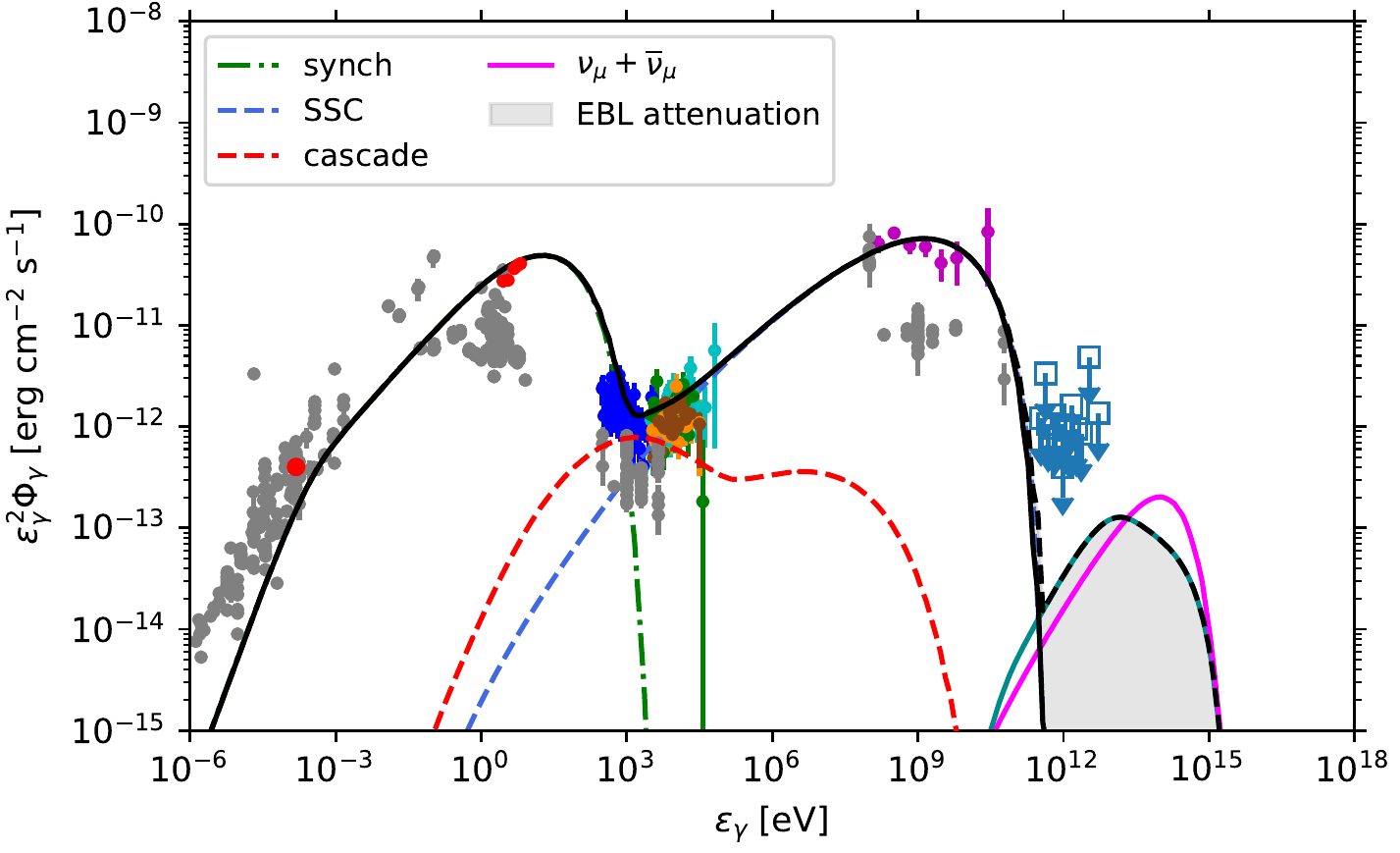}
\caption{Broadband SED model obtained for only synchrotron and synchrotron self-Compton emission with no external photon field. Red data points are UVOT, blue are XRT, green and orange are NuSTAR (FPMA \& FPMB), and magenta is GeV data from Fermi-LAT. The grey data points are the archival data. The left panel shows the purely leptonic emission case. Here the different line styles (shown in the legend for the synchrotron spectrum) correspond to the time evolution of the SED for a period of 10 days. The right panel shows the SED including $p\gamma$ interactions on the steady-state leptonic SED. The red dashed curve shows the secondary cascade radiation. The final spectrum is corrected for $\gamma\gamma$ pair production inside the jet and attenuation in the EBL (gray-shaded region). The magenta line is the muon neutrino flux obtained in this model for the best-fit proton maximum energy. The SEDs are produced for the marked period in Figure ~\ref{MWL} for MJD 59553-59563 (05/12/2021 to 15/12/2021). We also include the radio flux at 37 GHz shown in red color (\citealt{2021ATel15105....1K, Acharyya_2023}).}
\label{fig:hadronic}
\end{figure*}
\begin{figure*}
\centering
\includegraphics[width=0.49\textwidth]{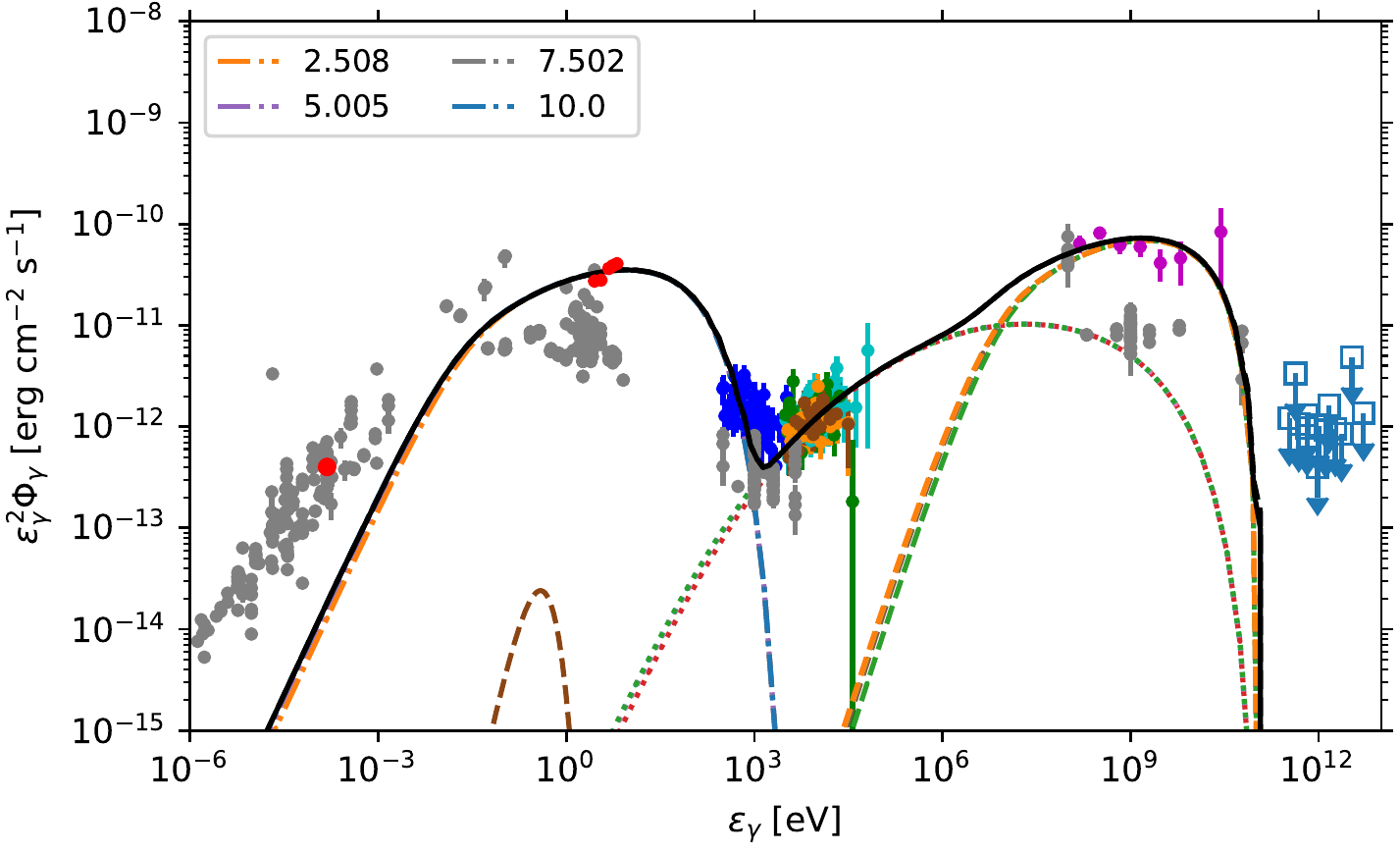}
\includegraphics[width=0.49\textwidth]{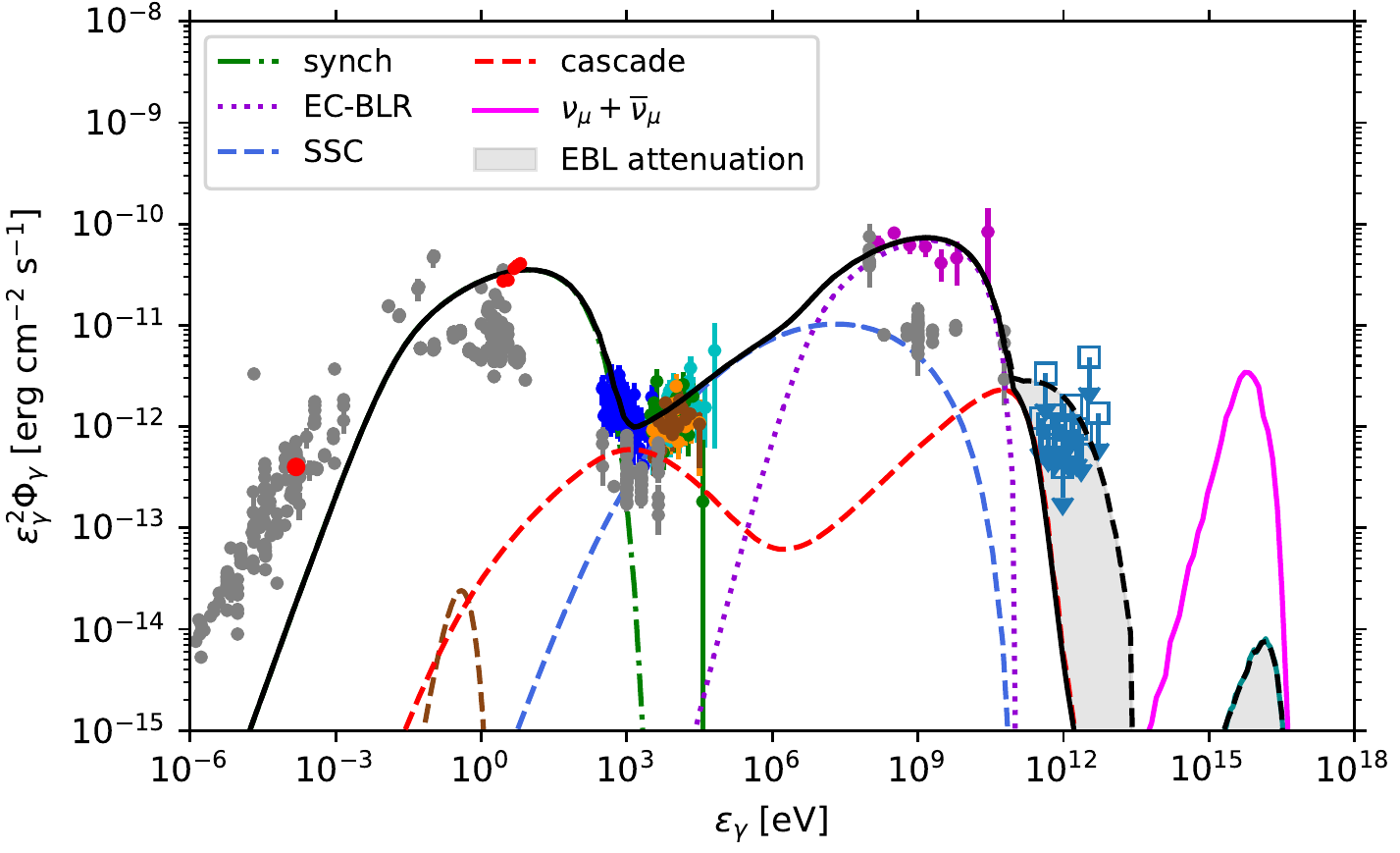}
\caption{Broadband SED modeled using synchrotron and inverse Compton emission off an external radiation field. The colors are the same as in Fig \ref{fig:hadronic}. The left panel shows the purely leptonic emission case. The different line styles (shown in the legend for the synchrotron spectrum) correspond to the time evolution of the SED over a period of 10 days. The right panel shows the SED including $p\gamma$ interactions on the steady-state leptonic SED and also on the external blackbody radiation field (dashed brown). The red dashed curve shows the secondary cascade radiation. The final spectrum is corrected for $\gamma\gamma$ pair production over all the available radiation fields in the jet emission region and attenuation in the EBL (gray-shaded region). The magenta line is the muon neutrino flux obtained in this model for the best-fit proton maximum energy. We also include the radio flux at 37 GHz (\citealt{2021ATel15105....1K}) shown in red square and the VHE data points from the VERITAS are shown for the comparison purpose in blue square \citep{Acharyya_2023}.}
\label{fig:with-External}
\end{figure*}

Broadband SED modeling is important to unveil the physical mechanism responsible for broadband emission and the simultaneous detection of the astrophysical neutrino by IceCube. From the temporal study, it is clear that during the time of neutrino detection, the source was in an active state in $\gamma$-ray, X-ray, and optical UV. We have combined the observations from all these wavebands and produced a broadband SED which we have modeled with a possible lepto-hadronic model.

We start with a pure leptonic scenario, where a distribution of leptons was injected inside the jet. The particle distribution evolved with time and later the synchrotron and synchrotron self-Compton (SSC) emission was derived.  We used a publicly available code, called \texttt{GAMERA}\footnote{\href{http://libgamera.github.io/GAMERA/docs/tutorials_main.html}{http://libgamera.github.io/GAMERA/docs/tutorials\_main.html}}, which solves the time-dependent transport equation given below,
\begin{equation}
\frac{\partial N(E,t)}{\partial t}= Q(E,t)-\frac{\partial}{\partial E}(b(E,t)N(E,t))-\frac{N(E,t)}{\tau\textsubscript{esc}}
\end{equation}
where, $Q(E,t)$ is the injected Lepton distribution and $b(E,t)$ corresponds to the energy loss rate by a synchrotron (SYN), SSC, and external Compton (EC) if there is an external radiation field and can be defined as $-{dE}/{dt}$. In the last term, we also included the constant escape time $\tau_{\rm esc} \sim R'/c$, where $R'$ denotes the size of the emitting region and $c$ is the speed of light. The escape of electrons over dynamical timescales $\tau_{\rm esc} \sim t_{\rm dyn} = R'/c$ is a prudent choice and has been widely considered in the literature.
This code uses the `Klein-Nishina' cross-section to compute Inverse Compton (IC) emission \citep{RevModPhys.42.237}.
 The escape time here represents the timescale over which adiabatic expansion losses limit the accumulation of relativistic electrons within the source emission region. So, the electrons that escape do not contribute to the observed emission (SYN/IC) any further and are rejected from the system. A quasi-steady state is reached in the emission region when the injection, $Q(E,t)$, is balanced by energy loss and/or escape, which is the ``cooling'' of the system. The time-dependent transport equation (Eqn.~3) is then used to calculate the electron energy distribution $dN/dE$ in this steady state, which subsequently gives the radiative spectra. 
Since the source is active simultaneously in all the wavebands and emitting neutrinos, we model the SED using a one-zone emission model. A log-parabola injection spectrum is assumed for electrons 
and the functional form of the LP distribution is given below,
\begin{equation}
Q(E)=L\textsubscript{0}(E/E_0)^{-(\alpha+\beta \log(E/E_0)}
\end{equation}
where $L_{0}$ is the normalization constant, $E_{0}$ is the reference energy, and $\alpha$, $\beta$ are the spectral index and the curvature parameter. 
 The important thing to note here is that the functional form of the particle spectrum (power law, log-parabola, power law with exponential cut-off, etc.) in jets of blazars is not known. Our choice of the particle spectrum is based on the convenience of fitting the data in our model. One can have a different particle distribution as well.

The cosmic ray protons accelerated inside the jet can interact with the soft photons produced in synchrotron radiation of leptons, thus undergoing various photohadronic processes. The timescale of these interactions can be expressed as 
\begin{equation}
\dfrac{1}{t'_{p{\rm \gamma}}} = \dfrac{c}{2\gamma_p'^2}\int_{\epsilon_{\rm th}/2\gamma_p}^\infty d\epsilon_\gamma'\dfrac{n(\epsilon'_\gamma)}{\epsilon_\gamma'^2}\int_{\epsilon_{th}}^{2\epsilon\gamma_p}d\epsilon_r \sigma(\epsilon_r) K(\epsilon_r)\epsilon_r \label{eqn:efficiency_ph}
,\end{equation}
where $\sigma(\epsilon_r)$ and $K(\epsilon_r)$ are respectively the cross-section and inelasticity of photopion production or BH pair production as a function of photon energy $\epsilon_r$ in the proton rest frame, and $n(\epsilon_\gamma')$ is the target photon number density \citep{Stecker:1968uc, Chodorowski_92, Mucke_00}. Neutrinos and high-energy photons are produced from the decay of charged and neutral pions produced in the photo-pion interactions ($p\gamma$ $\rightarrow$ $p + \pi^0$ or $n + \pi^+$). The decay of $\pi^+$ gives secondary leptons ($Q'_{e, \pi}$) and also results in a flux of $\nu_e$ and $\nu_\mu$. The high energy $\gamma$-rays are attenuated beyond TeV energies due to $\gamma\gamma$ pair production with the soft photons produced by the leptons (and also external photons for our external Compton model). As a result of these interactions, secondary $e^\pm$-s are produced. These secondary leptons along with those from $\pi^+$ decay and the $e^\pm$ pair produced in Bethe-Heitler (BH) interactions ($p\gamma$ $\rightarrow$ $p + e^{-}e^{+}$) generate secondary cascade emission. The spectrum of high-energy $\gamma$-rays from $\pi^\circ$ decay, the spectrum of $e^+$ and neutrinos from $\pi^+$ decay, and the spectrum of secondary $e^{-}e^{+}$ in BH process are estimated following the parametrization in  \citet{Kelner:2008ke}.  All the parameterizations and test cases presented there were reproduced, and then the proton and photon spectra were modified for our case. This method has been used earlier in many studies, including one of our earlier works \citep{Das:2022nyp}.

\begin{figure}
    \centering
    \includegraphics[width=0.47\textwidth]{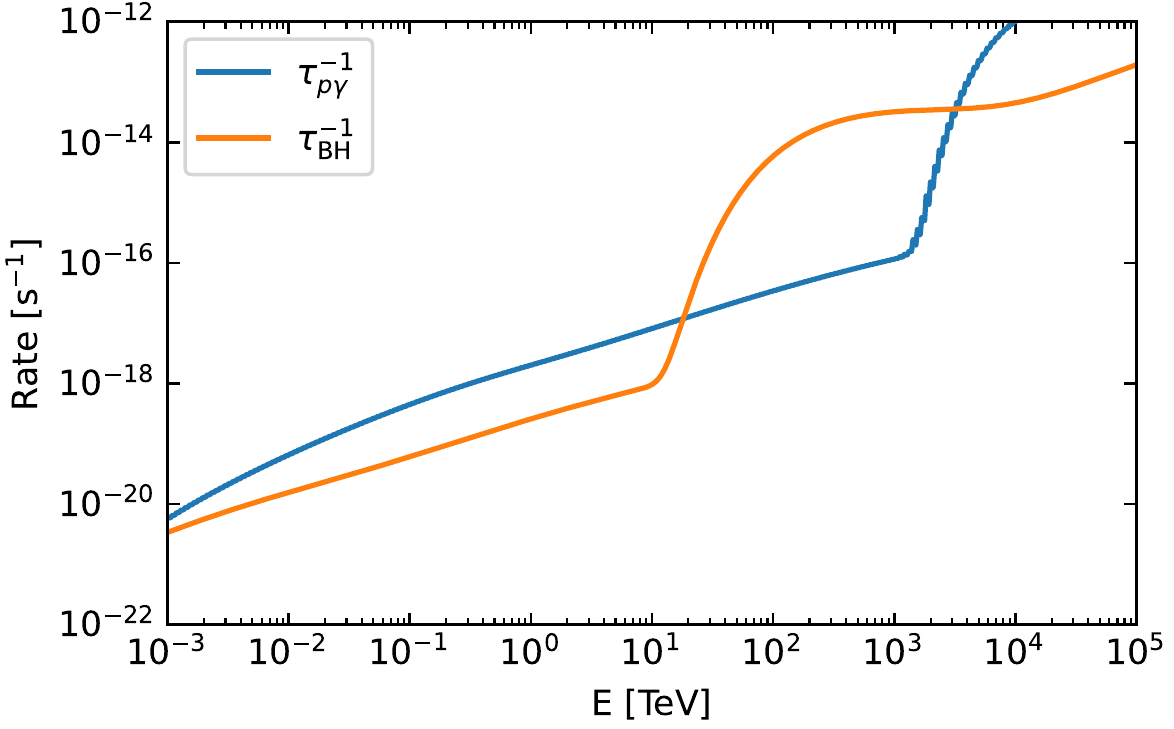}
    \caption{ A comparison of the rates of photopion interactions and Bethe-Heitler pair production processes.}
    \label{fig:op}
\end{figure}

 The high-energy electrons and positrons produced in $\gamma\gamma$ pair production ($Q'_{e,\gamma\gamma}$), charged pion decay ($Q'_{e,\pi}$), and BH process ($Q'_{e,\rm BH}$) can initiate cascade radiation from the jet.
We solve the steady-state spectrum of secondary electrons $N'_{e,s}(\gamma_e)$ in the 
jet frame using the analytical approach of \cite{Boettcher:2013wxa}, including $Q'_{e,\rm BH}$ in the source term. The escape term is to be the same as primary electrons. In a synchrotron-dominated cascade, emission from secondary electrons is given by $Q'_s(\epsilon'_s) = A_0 \epsilon'^{-3/2}_s \int_{1}^{\infty} d\gamma'_eN'_{e,s}(\gamma'_e)\gamma'^{-2/3}_e e^{-\epsilon'_s/b\gamma'^2_e}$. At such high energies, the emission of secondary leptons is synchrotron-dominated due to the suppression of IC efficiency by the Klein-Nishina effect. The flux of escaping high-energy $\gamma$-rays from pion decays is given as
\begin{equation}
    Q'_{\gamma, \rm esc}(\epsilon_\gamma') = Q'_{\gamma,\pi}(\epsilon'_\gamma) (1-\exp(-\tau_{\gamma\gamma}))/{\tau_{\gamma\gamma}}
\end{equation}
where, $\tau_{\gamma\gamma}$ is the absorption probability assuming an isotropic photon field, calculated using the formalism given in \cite{Gould_1967}.  To increase the efficiency of neutrino production, an escape timescale of protons longer than the photohadronic interaction timescale is required (see Fig.~\ref{fig:op}).  Otherwise, the required luminosity in protons increases drastically to obtain a substantial flux of neutrinos \citep[see, e.g.,][]{Das:2022nyp}. Hence, the energy loss due to proton escape is ignored. The relative weight of the rate of the corresponding process ($\pi$ production or BH interaction) is applied to the output fluxes. We assume a power-law distribution of protons, extending up to $E_{p, \rm max}$, given by $dN/dE_p\propto E_p^{-\alpha}$, where we consider the same injection spectral index $\alpha$ as for electrons. The normalization of the spectrum is fixed by the kinetic power required in protons to explain the observed SED. The muon neutrino flux at Earth is given as,
\begin{align}
    E_\nu^2J_\nu = \dfrac{1}{3} \dfrac{V'\delta_D^2\Gamma^2}{4\pi d_L^2}E'^2_\nu Q'_{\nu, p\gamma}
\end{align}
where $Q'_{\nu, p\gamma}$ is the total all-flavor neutrino flux produced in the co-moving jet frame. The factor 1/3 approximately takes into account the effect of neutrino oscillations. The $\nu_\mu+\overline{\nu}_\mu$ event rate at IceCube in a given operation time  $\Delta T$ is calculated using the expression
\begin{equation}
    \mathcal{N}_{\nu_\mu} =\Delta T \int_{\epsilon_{\nu, min}}^{\epsilon_{\nu,max}} d\epsilon_\nu \dfrac{d\Phi_\nu}{d\epsilon_\nu} \langle A_{\rm eff} \rangle_\theta
\end{equation}
where $\langle A_{\rm eff} \rangle_\theta$ is the effective detector area averaged over the zenith angle bin concerned \citep{IceCube:2018ndw}. We use the effective area for event selection as a function of neutrino energy from \cite{Stettner:2019tok}. The all-flavor neutrino flux can be approximated as three times the muon neutrino flux.

The IR-optical-UV and soft-X-ray emissions from PKS 0735+178 are expected to be produced by the synchrotron emission of relativistic leptons in a strong magnetic field. However, the hard X-ray and $\gamma$-ray are expected to be produced by the SSC process. In our analysis, although the X-ray data corresponds to 10 days around the time of flaring, we consider a steady proton injection, and hence the electromagnetic cascade and neutrino fluxes correspond to that produced in the flaring state once a steady state is reached as shown on the left panels of Fig.~\ref{fig:hadronic} and Fig.~\ref{fig:with-External}. The multiwavelength SED models obtained by using a pure leptonic and a lepto-hadronic emission model are shown in the left and right panels of Fig.~\ref{fig:hadronic}. The best-modeled parameters are shown in Table \ref{tab:sed_param}. 
The X-ray flux at keV energies provides stringent constraints on the secondary cascade emission and hence, limits the neutrino flux. We vary the proton maximum energy $E'_{p,max}$ in logarithmic intervals between 0.1 and 10 PeV to find the best model to SED and simultaneously maximize the neutrino flux. A steep cutoff in the proton spectrum is required to explain the observed neutrino event. Extending the proton spectrum to higher energies shifts the neutrino peak to higher energies. Hence a cutoff to the spectrum is required at $\gamma'_{p,\rm max}$ in our modeling. The neutrino flux peaks at $\sim0.1$ PeV. However, the luminosity requirement is very high in this case, much above the Eddington limit.

\subsection{With external photon field}

Since the source is argued to be a "masquerading BL Lac," it can also have an external radiation field, such as a broad line region and the dusty molecular torus. In the modeling, we consider that the emission region is within the boundary of the broad line region, and hence, the radiation field from BLR will dominate the emission in $\gamma$-ray and the neutrino event flux.
As argued in \citet{Ghisellini2011} the transition line between BL Lac and FSRQ type object can be given by the ratio of BLR luminosity to Eddington luminosity, $L_{\rm BLR}$/$L_{\rm Edd}$ $\sim$ 5$\times$10$^{-4}$.  Considering the black hole mass as $\sim 10^9 M_{\odot}$, the Eddington luminosity is estimated as 1.26$\times$10$^{47}$ erg/s which lead to the $L_{\rm BLR}$ = 6.3$\times$10$^{43}$ erg/s. Because a very small portion of the disk photon gets reprocessed in BLR (10$\%$), the disk luminosity can be derived as $L_{\rm disk}$ = 6.3$\times$10$^{44}$ erg/s. The BLR energy density in the co-moving frame can be estimated using the expression, 
\begin{equation}
    U'_{\rm BLR} = \frac{\Gamma^2 L_{\rm BLR}}{4 \pi R_{\rm BLR}^2 c}
\end{equation}
where, $R_{\rm BLR}$ $\sim$ 10$^{17}$ $L_{\rm disk, 45}^{0.5}$. Using the above values of $L_{\rm BLR}$ and $L_{\rm disk}$, the size of the broad line region and BLR energy density is estimated as, $R_{\rm BLR} = 0.79\times10^{17}$ cm and $U'_{\rm BLR} = 23.88 $ erg/cm$^3$, respectively. Please note that due to the uncertainties in the scaling relation, these values are correct within a factor of a few.
We also derive the location of the emission region down the jet using an expression, $d\sim 2 c \Gamma^2$ $t_{\rm var}/(1+z) \sim 2.33\times10^{18}$ cm, which suggests that the emission region is located at the outer boundary of the BLR. The BLR photon density sharply drops with increasing distance from the BLR \citep{10.1111/j.1365-2966.2009.15007.x} and therefore the estimated value for $U'_{\rm BLR}$ from the modeling is less than this value (Table \ref{tab:sed_param}). The size of the emission region in both cases is optimized during the modeling and is found to be less than what is expected from the variability estimation in eqn(\ref{eq:3}).


\begin{table}
\centering
\caption{ Multi-wavelength SED modeling results with the values of the best-model parameters. The Doppler factor and the Lorentz factor are fixed at 30.0 considering $\delta$ $\sim$ $\Gamma$. The redshift of the source is fixed at 0.45. }
 \begin{tabular}{l c c}
 \hline \noalign{\smallskip}
Parameters & SSC & SSC+EC \\
\noalign{\smallskip}  \hline  \noalign{\smallskip}
    $\delta_D$ & 30 & 30\\
    $B'$ [Gauss] & 0.55 & 4.5 \\
    $R'$ [cm] & $7.0\times 10^{15}$ & $2.0\times 10^{15}$\\
    $\alpha$ ($e/p$ spectral index) & 1.90 & 1.60 \\
    $\beta$ (log parabola curvature) & 0.04 & 0.001\\
    $E_0$ [GeV] & 0.09 & 0.1\\
    $\gamma'_{e,\rm min}$& 30 & 100 \\
    $\gamma'_{e,\rm max}$& 3.1$\times$10$^{4}$ & 1.0$\times$10$^{4}$\\
    $U'_{\rm BLR}$ [erg cm$^{-3}$]& -& 1.9 \\
    $T'_{\rm BLR}$ [K] & - & $5.0\times10^{4}$ \\
    $\gamma'_{p, \rm min}$  & 10.66 & 10.66 \\ 
    $\gamma'_{p, \rm max}$  & $1.7\times10^{5}$ & $4.26\times10^{6}$ \\
   \noalign{\smallskip}  \hline  \noalign{\smallskip}
    $L_{j,e}$ [erg s$^{-1}$] & $1.52\times10^{47}$ & $2.79\times 10^{44}$ \\
    $L_{j,B}$ [erg s$^{-1}$] & $0.1\times10^{45}$ & $6.8\times10^{45}$ \\
    $L_{j,p}$ [erg s$^{-1}$] & $6.3\times10^{49}$ & $6.3\times10^{47}$\\
    
 \noalign{\smallskip} \hline   \noalign{\smallskip}
 \end{tabular}
 \label{tab:sed_param}

\end{table}

\section{Discussions}

Neutrinos from astrophysical phenomenon is a smoking-gun evidence of cosmic-ray acceleration. The identification of the sources of IceCube detected diffuse astrophysical neutrino flux is of paramount importance to unveil extreme high-energy processes. Blazars are a potential candidate, among others, for neutrino emission through hadronic processes in their relativistic jets. Several sources have been identified in spatial coincidence with IceCube muon tracks. TXS~0506+056 is the only source associated spatially and temporally during the flaring activity in the Fermi-LAT waveband, with a $\nu_\mu$ event at the IceCube. However, the blazar PKS~0735+178 provides a testbed of one more such association. This BL Lac object lies within $90\%$ error region of one track-like event at IceCube of energy $E_\nu=172$ TeV. It can also be associated with a larger error region of a cascade event at the Baikal observatory of energy 43 TeV \citep{Dzhilkibaev_2022}, and an 18 TeV neutrino detected by KM3NeT \citep{Filippini_2022}.

In this article, we have studied the long-term $\gamma$-ray, X-ray, and UV optical data of the source. The SED reveals an active state during the IceCube detection epoch. The $\gamma$-ray and X-ray lightcurve exhibits a day-scale variability where the flux varies by a factor of $\sim3$ in the high state in comparison to the low state. 
In our one-zone lepto-hadronic modeling, we consider two cases viz., with and without external photon field. The absence of spectral emission lines suggests the source to be a BL Lac. However, the SSC spectrum provides insufficient target photons for neutrino production. The neutrino flux peaks at $\approx0.1$ PeV and the corresponding flux value is $\sim 2\times10^{-13}$ erg cm$^{2}$ s$^{-1}$. For higher values of $E'_{p, \rm max}$, the Bethe-Heitler cascade increases near the X-ray energies, hindering the increase in neutrino flux due to the restriction from Swift data. It should be noted, that the required kinetic power in protons exceeds the Eddington luminosity of the source by more than two orders of magnitude. We have checked that for protons to interact dominantly with the synchrotron photons, $E'_{p,\rm max}$ must be $\gtrsim 10^{17}$ eV, and hence $\Gamma E'_{p, \rm max}$ corresponds to ultrahigh energies, which may lead to signatures of the extragalactic cascade at multi-TeV energies \citep{Das:2021cdf}. 
In the absence of $\sim$TeV photon detection, such a scenario is possible but contentious.
Hence, for more reasonable results, PKS~0735+178 can be considered a ``masquerading BL Lac'' similar to TXS~0506+056, with hidden emission lines \citep{Padovani:2019xcv}. The external photons from the BLR region are then used as the primary target for $p\gamma$ interactions, producing a detectable neutrino flux.

 In our work, we consider an external photon field of temperature $5\times10^4$ K, i.e., within the optical frequency range. The X-ray data severely restricts the flux of synchrotron radiation from secondary electrons. The cascade flux in turn limits the allowed neutrino flux level. The neutrino event rate in our study is found to be 1.3$\times$10$^{-8}$ events per second. The expected events rate during the current high state of the source is $\sim0.12$ events per 100 days. Overall, the fit to the observed SED is significantly improved compared to the SSC model. A higher neutrino flux can be achieved if the peak of the external photon spectrum is further blueshifted. However, we have limited the photon field within the optical range, as considered widely in the literature for TXS~0506+056 \citep{Keivani:2018rnh}. Also, we emphasize the fact that an updated and detailed study of the X-ray flux and a fit to the archival radio data provide limited flexibility of the jet parameters. However, it should be noted that both the SSC model and the external Compton model require the proton spectrum to be cut off steeply at a specific energy, which is difficult to explain. One possibility is that, at higher energies, the cooling timescale may become shorter than the acceleration timescale or the acceleration process may become inefficient at higher energies reaching the limits in the physical size of the acceleration region.
 In a recent work by \cite{Acharyya_2023} they modeled the broadband SED including VERITAs and HESS SED data points with SSC, EC, and lepto-hadronic model. They predict the nominal neutrino flux at 170 TeV is 1.5 neutrinos per year (or 0.125 neutrinos per month) using an effective area of 30 m$^2$ which is a bit higher than our case.

 As shown in Table \ref{tab:sed_param}, the maximum energy of electrons and the protons differ very much which is because the electrons are rapidly cooled due to radiative losses and are difficult to accelerate to energies beyond sub-TeV energies. While protons are less efficient in radiation loss, and $\sim 2\times 10^3$ heavier than electrons, differing in the rest mass energy by the same factor. Hence, it will be more reasonable to compare their Lorentz factors and not energies for relativistic processes. From a modeling perspective, the energy transferred to neutrinos via the photopion interaction process is $E_\nu\approx E_p/20$. Hence to obtain a neutrino of energy $\sim 0.2$ PeV, the required proton maximum energy must be $ E_{p, \rm max} \approx 4$ PeV in the observer frame, which corresponds to $E'_{p, \rm max}=(1+z) E_{p, \rm max} /\Gamma \approx 0.2$ PeV in the comoving jet frame. However, the neutrino flux in our modeling peaks at a few PeV in the observer frame, requiring higher maximum proton energy. This is still within the energy uncertainty of the neutrino observation \citep{Sahakyan2023, Acharyya_2023}.

We find the X-ray -- neutrino correlation is more probable than the $\gamma$ -- neutrino correlation in the case of this source. A similar conclusion can be found for the TXS 0506+056 blazar in the literature \citep{Keivani:2018rnh}, in which case too the pion decay cascade is sub-dominant at the VHE $\gamma$-ray domain. However, the neutrino flux upper limit estimated by IceCube for 0.5-yr or 7.5-yr averaged observation of one event like IC-170922A is much higher than that obtained here. The IceCube detection potential is not sensitive to the flux level obtained here. Also, for the cascade event detection by Baikal, the angular uncertainty of positional correlation is poor, $\sim5.5^\circ$. In that case, one cannot ignore the possibility that other blazar sources within the $50\%$ containment zone can also be the possible source. However, for neutrino multiplets from this source, the estimated neutrino flux is more than an order of magnitude higher than that obtained here. Such high neutrino fluxes are extremely difficult to produce in a flare time of 21 days (comparing with \citealt{Sahakyan2023}), and one needs to invoke a hidden sector of neutrino production, or multiple zones, or neutrino production in the vicinity of the SMBH, outside the jet region.

The broadband SED of PKS 0735+178 has been modeled in a recent study by \citet{Sahakyan2023} where they have used a lepto-hadronic model and considered the three possible scenarios namely $p\gamma$ with proton-synchrotron, SSC and with external Compton. In $p$-synchrotron the model demands a very high magnetic field of the order of 100 Gauss, which is debatable in blazars. Their hybrid model with SSC required a very high proton luminosity exceeding the Eddington luminosity of the source. They also considered the external photon field to explain the high energy gamma-ray and eventually for the p-$\gamma$ interactions to explain the observed neutrino event. They found a neutrino event rate of 0.067 neutrinos per 21 days which corresponds to ~0.3 neutrinos in 100 days. However, the constraints from X-ray flux are not considered there. Their external photon field has a luminosity of $L_{\rm BLR}=2\times10^{43}$ erg/s, which peaks at $2\times10^{15}$ Hz (i.e. 8.3 eV).

 In our modeling, we noticed that the SSC model requires much higher jet power than the Eddington luminosity, which motivated us to include an external photon field. 
Even after including the external photon field the required jet power is higher than the Eddington luminosity. As mentioned earlier, in our model the external photon field peaks at the optical frequency and the jet luminosity required is $6.3 \times 10^{47}$ erg/sec, which is a factor of few higher than the Eddington luminosity of this source $L_{\rm Edd}=1.2\times 10^{46}(M_{\rm BH}/10^8 M_{\sun})$ erg/sec, where $M_{\sun}$ is the solar mass. The mass of the black hole $M_{BH}\sim 10^{8.8\pm 0.4} M_{\sun}$ has been assumed following \cite{Labita:2006jg}. 
Our estimated proton jet power is similar to \cite{Sahakyan2023}.
 The jet power in injected electrons obtained by other authors in blazar modeling varies within a range depending on the flare data and the model used to fit the data. The modeling done by \cite{Acharyya_2023} predicts a total jet power of 2.0$\times$10$^{48}$ erg/s, and a lower jet power is predicted for a harder proton spectrum, which is similar to our result.

\section{Conclusions}

The modeling of multiwavelength emission from jetted AGNs requires the knowledge of apparent jet speeds and their adiabatic evolution, the magnetic field, size and distance of emission region, etc. However, neutrino emission is a direct probe of cosmic-ray acceleration despite the degeneracy in these model parameters and depends dominantly on the available target photons for photopion production, providing essential constraints to blazar properties. In the case of BL Lac objects, neutrino production is suppressed in a one-zone photohadronic model due to the lack of sufficient target photons from SSC. In this work, we have studied a recent association of $\gamma$-ray blazar PKS~0735+178 with an IceCube muon track detected on Dec 2021, without invoking any exotic physical phenomenon. We predict the resulting neutrino fluxes, aided by a multiwavelength SED model obtained in various scenarios, with and without an external blackbody radiation field. The obtained flux levels are still much lower than the IceCube operation sensitivity. Resolving the uncertainty around the temperature and density of these external seed photons for photohadronic processes is a non-trivial task, and we restrict ourselves to optimal values found in the literature. Interestingly, an association between the X-ray flux and neutrino emission is irrefutable. We conclude that the blazar's contribution to the diffuse neutrino background cannot be neglected. Upcoming detectors with improved sensitivity and angular resolution will be able to pinpoint the neutrino-emitting blazars and constrain our neutrino flux predictions.

\section*{Acknowledgements}
 We thank the anonymous referee for insightful comments and suggestions which have helped to improve the scientific clarity of our work.
R. Prince and B.Z. are grateful for the support of the Polish Funding Agency National Science Centre, project 2017/26/A/ST9/-00756 (MAESTRO 9) and the European Research Council (ERC) under the European Union’s Horizon 2020 research and innovation program (grant agreement No. [951549]).
The work of S.D. was supported by JSPS KAKENHI Grant Number 20H05852. Numerical computation in this work was partly carried out at the Yukawa Institute Computer Facility.

\section*{Data Availability}

All the data are publicly available and can be accessed easily.



\bibliographystyle{mnras}
\bibliography{mnras} 




\appendix




\label{lastpage}
\end{document}